\documentclass[preview]{elsarticle}

\def	\be	{\begin{equation}}
\def	\ee	{\end{equation}}
\def	\bqt	{\begin{quote}}
\def	\eqt	{\end{quote}}

\def	\lf	{\left (}
\def	\rt	{\right )}

\def	\comma	{\quad , \quad}

\begin{document}

\title{Logarithmic Corrections to Gravitational Entropy and the Null Energy Condition}

\author{Maulik Parikh}

\author{Andrew Svesko}

\address{Department of Physics and Beyond: Center for Fundamental Concepts in Science \\
Arizona State University, Tempe, Arizona 85287, USA}

\begin{abstract}
Using a relation between the thermodynamics of local horizons and the null energy condition, we consider the effects of quantum corrections to the gravitational entropy. In particular, we find that the geometric form of the null energy condition is not affected by the inclusion of logarithmic corrections to the Bekenstein-Hawking entropy. 
\end{abstract}

\maketitle

\section{Introduction}

The null energy condition (NEC) requires that, for any null vector $v^\mu$ at any point in spacetime, 
\be
T_{\mu\nu} v^\mu v^\nu \geq 0 \; .	\label{TNEC}
\ee
This condition plays a critical role in classical general relativity, being used in proofs of the singularity theorems \cite{hawkingpenrose} as well as in black hole thermodynamics \cite{bardeen}.
Expressed in the form of (\ref{TNEC}), the NEC appears as a property of matter, since it is defined in terms of the matter energy-momentum tensor. But even in some classical limit,  quantum field theory, our best framework for describing matter, does not appear to have a consistency requirement of the form of (\ref{TNEC}), at least when matter is considered in isolation without coupling to gravity.

A clue to the origin of the null energy condition comes from its role in general relativity. Instead of being a property purely of matter, perhaps the null energy condition is really a property of a combined theory of matter and gravity. In this context, Einstein's equations imply a different, though equivalent, form of the NEC,
\be
R_{\mu\nu} v^\mu v^\nu \geq 0 \; ,	\label{RNEC}
\ee
where $R_{\mu\nu}$ is the Ricci tensor. Written in this way, the NEC can be interpreted as a constraint on spacetime geometry, rather than as a constraint on matter. Indeed, it is this geometric form of the null energy condition, known as the Ricci or null convergence condition, that is ultimately invoked in gravitational theorems, because it is the Ricci tensor that appears in the  Raychaudhuri equation. Thus, if (\ref{RNEC}) could be derived directly, we could uphold all the gravitational theorems that otherwise rely on the so-far-unprovable (\ref{TNEC}).

Recently, it has been shown that precisely this condition can be derived from string theory \cite{NECderivation}, which, indeed, is a combined theory of matter and gravity; the NEC arises essentially as the spacetime interpretation of the Virasoro constraint on the worldsheet. But remarkably, the NEC in its geometric form, (\ref{RNEC}), can also be derived in a completely different way: as a consequence of the second law of thermodynamics \cite{thermoNEC}. Specifically, one assumes there exists some underlying microscopic theory obeying the laws of thermodynamics, from which classical gravity emerges via some coarse-graining procedure. 

The notion that gravity is emergent is no longer controversial. For example, in the AdS/CFT correspondence, quantum gravity in anti-de Sitter space is described by a conformal field theory dual. The conformal field theory does not itself contain the graviton among its fundamental degrees of freedom so, from the dual point of view, gravity as well as an extra spatial dimension are emergent phenomena. In this approach, gravity emerges globally. However, in a remarkable paper \cite{einsteineqnofstate}, Jacobson considered a local version of holography. The idea was to assume, in keeping with the universality of horizon entropy, that gravitational entropy can be associated to ``local Rindler horizons". Such local horizons exist everywhere because, in the vicinity of any point, spacetime is effectively flat and the local Minkowski space can be expressed in accelerating coordinates. Jacobson then found that
Einstein's equations arise from the first law of thermodynamics, applied to a local Rindler horizon. In the same spirit, as shown in \cite{thermoNEC}, the null energy condition, in the form of (\ref{RNEC}), arises from the second law of thermodynamics, applied locally.  

The string and emergent gravity derivations both consider classical matter and gravity. The natural next question is to ask whether quantum effects lead to violations of the NEC. Indeed, it is known that the matter form of the NEC is violated when first order quantum effects are taken into account, e.g., by Casimir energy. Nevertheless, it is not clear that this indicates a violation in the Ricci convergence condition  (\ref{RNEC}). To understand this, consider the semi-classical Einstein equations,
\be
G_{\mu\nu}=8\pi G\langle T_{\mu\nu}\rangle\;, \label{semi} 
\ee
which describe the backreaction of quantum fields on a classical background. The effect of the fluctuating quantum fields is captured by the renormalized expectation value of the energy-momentum tensor $\langle T_{\mu\nu}\rangle$ over a particular background. The relevance of  $\langle T_{\mu \nu} \rangle$ to spacetime geometry relies on the validity of an equation of the form of (\ref{semi}), but we are not aware of any rigorous derivation of this equation as the semi-classical limit of a theory of both quantum matter and quantum geometry. Indeed, an equation which treats gravity classically but matter quantum-mechanically appears to be in some tension with the spirit of string theory in which matter and gravity are treated in a unified manner. In principle $\langle T_{\mu\nu}\rangle$ can be derived from an effective action $S_{\rm eff}(g_{\mu\nu})$ describing the quantum matter fields propagating on the background metric $g_{\mu\nu}$. In that case, generally one finds that $\langle T_{\mu\nu}\rangle$ will depend on higher-curvature terms (see, e.g., \cite{birrelldavies}). The field equations, therefore, will in general include higher-curvature corrections to Einstein's equations, severing the link between the NEC as a constraint on matter  (\ref{TNEC}) and the NEC as a constraint on geometry  (\ref{RNEC}). Thus a violation in  (\ref{TNEC}) does not imply a violation in  (\ref{RNEC}), and vice versa. 

In this note, we take a different approach. Rather than calculating $\langle T_{\mu \nu} \rangle$, and then trying to determine its gravitational implications, the novel idea here is to directly determine $R_{\mu \nu} v^\mu v^\nu$ in the semi-classical theory. Specifically, we use the known form of the quantum-corrected version of the Bekenstein-Hawking entropy \cite{Kaul00-1} to obtain the Ricci convergence condition. We find that, if we replace the  Bekenstein-Hawking entropy of a horizon with its one-loop generalization and apply the second law of thermodynamics, we again arrive at exactly the Ricci convergence condition (\ref{RNEC}). Quantum corrections, at least of the type that contribute to the entropy, do not appear to alter the condition; if these were the only quantum corrections, then, for example, singularity theorems would continue to hold even in the semi-classical theory.


\section{Time Derivatives of Entropy}
\indent

In the emergent gravity paradigm, gravity emerges out of the coarse-graining of some more fundamental microscopic system. Holography suggests that the degrees of freedom of this system presumably live in one dimension less than the dimensionality of spacetime. In particular, we will assume that classical gravity in spacetime emerges in some suitable thermodynamic limit of the underlying system, such that the coarse-grained entropy of the dual system corresponds to the local Bekenstein-Hawking entropy of a null congruence, as defined in the next section. (We emphasize that the thermodynamic matter system we will have in mind will always be this dual system rather than any matter that may live within spacetime.) Little is known about this microscopic theory. Nevertheless, a few general remarks can be made. Here we will review the discussion  in \cite{thermoNEC}. Consider a finite thermodynamic system and let $S_{\rm max}$ be its maximum coarse-grained entropy. Broadly, there are two kinds of thermal systems: those that are at thermodynamic equilibrium, and those that are approaching equilibrium. For systems already at equilibrium, $S = S_{\rm max}$, and $\dot{S}, \ddot{S} = 0$.

For systems approaching equilibrium, $S < S_{\rm max}$ and the second law says that $\dot{S} \geq 0$. We will also be interested in the second derivative of the entropy. Now, since the entropy tends to a finite maximum value as it approaches thermal equilibirum, and since $\dot{S}\geq0$, it seems intuitively reasonable that the time derivative of entropy will be a decreasing function of time, so that $\ddot{S}\leq0$. This inequality holds for a great many systems of interest. For example, consider a gas diffusing in 3+1 dimensions. Starting with an initial Gaussian density profile $\rho (r, 0) \sim e^{-r^2/2}$, the diffusion equation gives the density profile at later times:
\be
\rho(r,\tau)=\frac{1}{2\left(\pi\left(1+2D\tau\right)\right)^{3/2}}\exp\left(-\frac{r^{2}}{2\left(1+2D\tau\right)}\right) \; ,
\ee
where $D$ is the diffusion constant.
A straightforward calculation then yields the entropy:
\be
S(\tau)=-\int dV\rho\ln\rho \sim \frac{3}{2} \ln(1+2D\tau) \; .
\ee
It is easy to verify that $\ddot{S}=-(2/3)\dot{S}^{2}$, from which we see that 
\be
S \geq0,\quad\dot{S}\geq0,\quad\ddot{S}\leq0 \; , \label{entcond}
\ee
at all times. 

As another example, consider a system of $N$ macrostates. Since classical phase space trajectories are typically chaotic, the dynamics can be described probabilistically after a few Lyapunov times. Assuming the ergodic hypothesis, the probability distribution can be taken to be uniform over all microstates. Then a short argument  \cite{thermoNEC} yields
\be
S(t)\approx S_{\rm max}(1-e^{-kt}) \; ,
\ee
where $k$ is some constant with units of inverse time, and $S_{\rm max}$ is the maximum entropy of the system, corresponding to the largest macrostate. This form of the entropy also satisfies (\ref{entcond}) for all times. In particular, $\ddot{S} < 0$.

In fact, the non-positivity of $\ddot{S}$ can be proven quite generally \cite{falkovitch} using Onsager theory, at least for a very broad class of near-equilibrium systems approaching equilibrium. Near equilibrium, the time-derivative of a density fluctuation obeys a linear Onsager relation:
\be
\delta \dot{\rho} = \hat{L} \delta \rho \; .
\ee
Onsager argued on rather general grounds that whenever certain time-reversibility properties are satisfied, the matrix $\hat{L}$ is symmetric \cite{onsager}. From this one can show  \cite{falkovitch} that $\ddot{S} \leq 0$. That is, when $\hat{L}$ is symmetric, as is generally the case, then a sufficient condition for the non-positivity of $\ddot{S}$ is that the system be near equilibrium; for such systems (\ref{entcond}) apply. This is a sufficient condition but, as the examples above illustrate, not a necessary one; $\ddot{S}$ can even be non-positive away from equilibrium. Henceforth, we will assume that conditions (\ref{entcond}) are indeed satisfied by the microscopic dual theory of gravity.


\section{Thermodynamic Origin of the Null Energy Condition}
\indent

Let us now translate these conditions on entropy to spacetime. The idea in the emergent gravity paradigm is to associate an underlying non-gravitational system, presumably living in one dimension less, such that the coarse-grained entropy of the the microscopic system accounts for the Bekenstein-Hawking entropy. Now, the universality of the Bekenstein-Hawking entropy formula, derived originally for black hole horizons, suggests that it  really counts gravitational degrees of freedom. This led Jacobson to associate gravitational entropy {\em locally} to local Rindler horizons, thereby obtaining Einstein's equations from the Clausius relation. In a similar vein, we shall assume here that we can associate a dual thermodynamic system to every non-contracting infinitesimal patch of every future-directed null congruence at every point in spacetime. The requirement that the patch be non-contracting towards the future ensures that its gravitational entropy will be non-decreasing, as would be expected for an underlying system obeying the second law of thermodynamics. Consider some such patch.
Then
\be
S = \frac{A}{4} \; . \label{BHent}
\ee
Use of the Bekenstein-Hawking entropy formula amounts to an implicit assumption that classical physics is described by Einstein gravity minimally coupled to matter; for higher-curvature theories of gravity, or for non-minimally coupled gravity, the Bekenstein-Hawking entropy would have to be replaced by its appropriate generalization, such as the Wald entropy \cite{waldentropy}. Later we will consider one-loop corrections to (\ref{BHent}). Identifying time in the thermodynamic system with the affine parameter, $\lambda$, of the null congruence, we find
\be
\dot{S} = \frac{A}{4} \theta \comma \ddot{S} = \frac{A}{4} \lf \theta^2 + \dot{\theta} \rt \; , \label{Sdots}
\ee
where $\theta = d \ln A/d \lambda$ is the congruence expansion parameter. $\theta$ is roughly constant over the surface; this can always be arranged by making the surface small enough. Because the congruence is light-like, its generators obey the optical Raychaudhuri equation:
\be
\dot{\theta}= - \frac{1}{2} \theta^2 - \sigma^2 + \omega^2 - R_{\mu\nu} v^\mu v^\nu \; .
\ee
By hypersurface-orthogonality, $\omega^2 = 0$. We can also choose a surface for which the shear is initially zero. Then, for small times, $\sigma^2$ is negligible. We will not need to assume that $\theta$ is small though, if the system is near equilibrium, the $\theta^2$ term is negligible. 

Dropping $\sigma^{2}$ and $\omega^{2}$, let us consider the Raychaudhuri equation separately for the two types of thermodyamic systems. For systems already at equilibrium, $\dot{S}$ and $\ddot{S}$ are both zero and so, from (\ref{Sdots}), $\theta$ and $\dot{\theta}$ are also zero. (We emphasize again that the system considered to be at equilibrium here is the dual theory of matter whose coarse-grained entropy is the gravitational entropy.) We therefore have
\be
R_{\mu \nu} v^\mu v^\nu = 0 \; . \label{RNECsat}
\ee
Next, consider systems approaching equilibrium. Then $\dot{S} > 0$ and $\ddot{S} \leq 0$. Correspondingly, $\theta > 0$ and $\dot{\theta} + \theta^2 \leq 0$. Then we have 
\begin{eqnarray}
R_{\mu \nu} v^\mu v^\nu & = & - ( \dot{\theta} + \theta^2 ) + \frac{1}{2} \theta^2 \nonumber \\
& = & - \frac{\ddot{S}}{S} + \frac{1}{2} \left (\frac{\dot{S}}{S} \right)^{\! \! 2} \nonumber \\
& > & 0 \; . \label{RNECexp}
\end{eqnarray}
We see that for congruences corresponding to equilibrium systems, (\ref{RNECsat}) holds, while for those congruences corresponding to non-equilibrium systems, (\ref{RNECexp}) holds. In either case, thermodynamics implies that $R_{\mu \nu} v^\mu v^\nu \geq 0$, which is precisely the geometric form of the null energy condition, (\ref{RNEC}).

To complete the proof, we need to show that, for any spacetime point $p$, and any future-directed null vector $v$ in the tangent space at $p$, there exists a null congruence containing an integral curve through $p$ whose tangent at $p$ is $v$, and which corresponds to a thermodynamic system of either type. We show this by construction. If, for a given $v$, $R_{\mu \nu} v^\mu v^\nu$ vanishes at $p$, then we are dealing with an equilibrium system with $\theta, \dot{\theta} = 0$; the corresponding congruence could be a piece of a local Rindler horizon containing $v$. Alternatively, if $R_{\mu \nu} v^\mu v^\nu > 0$ at $p$ for the given $v$, then the  thermodynamic system is not at equilibrium. An example of the corresponding congruence might be a piece of the integral curves of the future light cone of some earlier point $p'$. This establishes the null energy condition.

This is a remarkable result. The classical null energy condition, a condition that in the form of (\ref{TNEC}) has proven impossible to derive from quantum field theory, emerges quite naturally in the form of (\ref{RNEC}) from the local thermodynamics of gravity. One could regard this as additional evidence that gravity is emergent, which was the premise of the calculation.

An obvious next question to ask is whether the result can be extended to the quantum regime. Much effort has been put into calculating quantum corrections to the NEC on the matter side \cite{grahamolum,kontou,QNEC}. Fortunately, there is an easier way to address this question. The key point is that the Raychaudhuri equation depends only on the geometry of spacetime and not on the theory in which the geometry arises. In particular, it should hold also for the geometry that arises in an effective theory of gravity that includes one-loop corrections. Furthermore, the Raychaudhuri equation contains the actual geometric object of interest, namely $R_{\mu \nu} v^\mu v^\nu$. It is the positivity of this term that controls the possible existence of singularities, say. By contrast, the gravitational implications of $\langle T_{\mu \nu} \rangle$ rely on the unclear question of how quantum matter couples to gravity. If, for example, the left-hand side of Einstein's equations are modified by the inclusion of geometric counter-terms, then the sign of $\langle T_{\mu \nu} \rangle v^\mu v^\nu$ does not have any obvious bearing on the sign of $R_{\mu \nu} v^\mu v^\nu$.

The sign of $R_{\mu \nu} v^\mu v^\nu$ is determined by the Raychaudhuri equation once we know $\theta, \dot{\theta}$. Our underlying (and non-trivial) assumption is that the semi-classical theory can continue to be described by thermodynamics. Under that assumption, we need to express geometric quantities like $\theta, \dot{\theta}$ in terms of thermodynamic quantities, specifically time derivatives of the coarse-grained entropy. The one-loop quantum-corrected formula for the gravitational entropy is
\be 
S=\frac{A}{4}+c\ln A+\mathcal{O}(1)\;,
\ee
where $c$ is a constant. As before we have suppressed Newton's constant here, so that $A$ is measured in Planck units. Such a logarithmic correction \cite{Kaul00-1} to the Bekenstein-Hawking entropy arises in a great variety of contexts. These include Carlip's derivation using the Virasoro algebra associated with two-dimensional conformal symmetry at the horizon \cite{Carlip00-1}, the partition function of the BTZ black hole \cite{Govindarajan01-1}, one-loop effects  \cite{Fursaev95-1,Mann98-1}, type-A (Euler density) contribution to the trace-anomaly induced effective action \cite{Cai09-1,Aros10-1}, along with many others; see, e.g. \cite{Medved05-1,Page05-1} for a review. The technical reason for this evident universality of the leading correction to the Bekenstein-Hawking entropy is that all the microscopic derivations ultimately invoke the Cardy formula. 

Note that the positivity of $S$ implies that
\be
c \geq -\frac{A}{4 \ln A} \; . \label{boundonc}
\ee
Typically, $c$ is of order unity. In fact, the majority of calculations agree that
\be
c = -\frac{3}{2} \; , \label{c=-3/2}
\ee
with some other approaches giving a result that differs by a factor of order unity. As $A$ is measured in Planck units, the validity of an approximately classical regime requires that $A \gg 1$. Then we have from (\ref{boundonc}) that $\left(\frac{A}{4} + c \right) > 0$.

The time derivative of the entropy is given by
\be
\dot{S}= \theta \left ( \frac{A\theta}{4} + c \right ) \; .
\ee
Hence $\dot{S} \geq 0 \Rightarrow \theta \geq 0$: increasing entropy corresponds to expanding congruences, unsurprisingly. The second derivative of the entropy is
\be
\ddot{S}=\frac{A}{4}\left[\left(\frac{A}{4} + c\right)\dot{\theta}+\theta^{2}\right]\;.
\ee
Next, as we are regarding the gravitational entropy to be the coarse-grained entropy of some dual thermodynamic system, we invert the geometric quantites $A$, $\theta$, and $\dot{\theta}$ in terms of the thermodynamic quantities $S$, $\dot{S}$, and $\ddot{S}$. We find
\be
A(S) = 4c \, W \! \left ( \frac{e^{S/c}}{4c} \right ) \; ,
\ee
where $W$ is the Lambert W-function, and
\be
\theta = \frac{\dot{S}}{\frac{A(S)}{4} + c} \comma \dot{\theta} = \frac{\ddot{S}}{\frac{A(S)}{4} + c}  -\frac{\dot{S}^2 A(S)}{4} \left ( \frac{1}{\frac{A(S)}{4} + c} \right )^{\! 3} \; .
\ee
We can now again consider the two types of thermal systems. For systems at equilibrium we  have $\dot{S}=\ddot{S}=0$, so that $\theta=\dot{\theta}=0$, leading to (\ref{RNECsat}) via the Raychaudhuri equation. For systems approaching equilibrium we find, using (\ref{entcond}), that
\begin{eqnarray}
R_{\mu \nu} v^\mu v^\nu & = & - \dot{\theta} - \frac{1}{2} \theta^2 \nonumber \\
& = & \frac{1}{\frac{A(S)}{4} + c} \left [ - \ddot{S} + \frac{1}{2} \left ( \frac{\dot{S}}{\frac{A(S)}{4} + c} \right )^{\! 2} \left( \frac{A(S)}{4} - c \right ) \right ] \nonumber \\
& >& 0 \; ,
\end{eqnarray}
provided $c< A/4$. This is indeed the case since $A \gg 1$ and explicit calculations indicate that $c$ is of order unity, (\ref{c=-3/2}). Therefore, even in the context of semi-classical gravity, we again recover the geometric form of the null energy condition from the second law of thermodynamics.


\section{Discussion}

The null energy condition has traditionally been regarded as a property of matter. Recently we have postulated that the NEC arises from a more fundamental theory combining matter and gravity \cite{NECderivation,thermoNEC,tworoads}. In particular, by identifying the coarse-grained entropy of this fundamental, underlying theory with the gravitational entropy, we were able to explicitly derive the geometric form of the NEC, the Ricci convergence condition, from the second law of thermodynamics. Here, we have taken a novel approach to studying quantum effects in semi-classical gravity. In particular, we have shown that the Ricci convergence condition remains stable under one-loop quantum corrections to the Bekenstein-Hawking entropy. If this were the entirety of the effect (which we do not claim), it would mean that quantum effects at one-loop do not, for example, prevent the occurence of cosmological or black hole singularities. 

There are at least two clear instances of quantum effects violating the matter form of the null energy condition: Hawking radiation and Casimir energy. Hawking radiation, however, is really a non-perturbative effect; this is easiest to understand by noting that Hawking radiation can be expressed as a tunneling process \cite{tunneling,secret}. But Casimir energy certainly violates the matter NEC at one-loop. How is our result to be reconciled with the general expectation that the matter null energy condition should be violated by one-loop effects? Here it is important to recognize that it is not definitively known how Casimir energy actually gravitates. One can imagine several possibilities. Since quantum corrections inevitably induce gravitational counter-terms, these would generically sever the link between the matter and the geometry form of the NEC. Thus it could be that the matter NEC is indeed violated by one-loop quantum effects, but the geometric one is not. Alternatively, it could be that vacuum expectation values of $T_{\mu \nu}$ do not gravitate for unknown reasons related to the resolution of the cosmological constant problem. Or it could be that there are additional quantum gravity effects  that are not captured by the logarithmic correction to the entropy considered here. Finally, it could be that only classical spacetime physics corresponds to thermodynamics in the dual theory, and that the approach here is invalid. It would be interesting to study these issues further.


\end{document}